\documentclass[twocolumn]{autart}

\usepackage{graphicx}
\usepackage{cite}
\usepackage{subfigure}
\usepackage{array}
\usepackage{color}
\usepackage{amsmath, amsfonts,amssymb}
\usepackage{float}
\usepackage{CJK}
\usepackage{graphicx,epstopdf}
\usepackage{indentfirst}
\usepackage{bm}
\usepackage{mathbbold}
\usepackage{mathrsfs}
\usepackage{float}
\usepackage{ccaption}
\usepackage[caption=false,font=normalsize,labelfont=sf,textfont=sf]{subfig}
\usepackage{enumerate}

\newtheorem{remark}{Remark}[section]
\newtheorem{lemma}{Lemma}[section]

\newtheorem{corollary}{Corollary}[section]
\newtheorem{example}{Example}[section]

\newcommand\rsx[1]{\left.{#1}\vphantom{\Big|}\right|}

\def\ba{\begin{array}}
\def\ea{\end{array}}
\def\ban{\begin{eqnarray*}}
\def\ean{\end{eqnarray*}}
\def\be{\begin{equation}}
\def\ee{\end{equation}}
\def\bna{\begin{eqnarray}}
\def\ena{\end{eqnarray}}

\def\nn{\nonumber}

\def\dref#1{(\ref{#1})}

\def\QEDclosed{\mbox{\rule[0pt]{1.3ex}{1.3ex}}}

\def\QED{\QEDclosed} % default to closed

\def\proof{\noindent\hspace{2em}{\itshape Proof: }}
\def\endproof{\hspace*{\fill}~\QED\par\endtrivlist\unskip}

\def\a{\alpha}
\def\b{\beta}

\def\d{\delta}

\def\g{\gamma}

\def\P{\Phi}
\def\S{\Psi}

\def\s{\sigma}

\def\x{\xi}

\begin{document}

\begin{frontmatter}

%\runtitle{Controllability of networked MIMO systems}

\title{Controllability of networked MIMO systems\thanksref{footnoteinfo}}

\thanks[footnoteinfo]{Corresponding author X. F. Wang. Tel.:
+86-21-5474-7511. Fax: +86-21-5474-7161.}

\author[sjtu]{Lin Wang}\ead{wanglin@sjtu.edu.cn},
\author[cityu]{Guanrong Chen}\ead{eegchen@cityu.edu.hk},
\author[sjtu]{Xiaofan Wang}\ead{xfwang@sjtu.edu.cn},
\author[cityu]{Wallace K. S. Tang}\ead{eekstang@cityu.edu.hk}

\address[sjtu]{Department of Automation, Shanghai Jiao Tong University,
and Key Laboratory of System Control and Information Processing,
Ministry of Education, Shanghai, China}
\address[cityu]{Department of Electronic Engineering, City University
of Hong Kong, Hong Kong, China}


\begin{abstract}
In this paper, we consider the state controllability of networked
systems, where the network topology is directed and weighted and the
nodes are higher-dimensional linear time-invariant (LTI) dynamical
systems. We investigate how the network topology, the node-system
dynamics, the external control inputs, and the inner interactions
affect the controllability of a networked system, and show that for
a general networked multi-input/multi-output (MIMO) system: 1) the
controllability of the overall network is an integrated result of
the aforementioned relevant factors, which cannot be decoupled into
the controllability of individual node-systems and the properties
solely determined by the network topology, quite different from the
familiar notion of consensus or formation controllability; 2) if the
network topology is uncontrollable by external inputs, then the
networked system with identical nodes will be uncontrollable, even
if it is structurally controllable; 3) with a controllable network
topology, controllability and observability of the nodes together
are necessary for the controllability of the networked systems under
some mild conditions, but nevertheless they are not sufficient. For
a networked system with single-input/single-output (SISO) LTI nodes,
we present precise necessary and sufficient conditions for the
controllability of a general network topology.
\end{abstract}

\end{frontmatter}

\section{Introduction}

Complex networks of dynamical systems are ubiquitous in nature and
science, as well as in engineering and technology. When control to a
network is taken into consideration, the controllability of the
network is essential, which is a classical concept \cite{Kalman}
applicable to multi-variable control systems \cite{Gilbert,
Shields}, composite systems \cite{davison} and decentralized control
systems \cite{Kobayashi, Tarokh}, etc.

The subject of system controllability has been extensively studied
over the last half a century. To date, various criteria have been
well developed, including different kinds of matrix rank conditions
and substantial graphic properties \cite{Gilbert, Hautus, Lin74,
davison, Glover, Shields, Kobayashi, Tarokh, Jarc, yyliu, RR2014}.
One closely-related subject is the controllability of multi-agent
systems, including the controllability of consensus systems
\cite{bliu, Rahmani, Lou, ji, Zhang, Nabi, Ni2013, Xiang}, formation
\cite{Cai}, and pinning strategies \cite{chen}.

%In all the above-mentioned investigations, in addition to linear
%algebra, graph theory is a typical tool to use. Graph theory has
%been extensively studied since the time of Euler, where a graph is
%a representation of a set of nodes interconnected by edges. When
%applied to physical networks, each node represents a dynamical
%system and each edge represents a communication channel between
%two connected node-systems, where the node can be
%higher-dimensional and the edge can be directed, weighted and
%multi-dimensional \cite{chenwangli}.

%When applied to complex networks, the controllability issue
%becomes more complicated and challenging, for which some efficient
%conditions for large-scale networks have been developed \cite{Cowan,
%Wang, WangB, Yan, Gao, Meni, Wang14, Yuan}. For example, maximum
%matching is used to find driver nodes to guarantee the structural
%controllability \cite{yyliu, Chapman}, and integer programming is
%used to find the maximum controllable subspace \cite{Liu12}. It is
%noted that most of these algorithms are only applicable to the
%study of structural controllability, but may not work for state
%controllability after all.

Noticeably, many existing results on controllability are derived
under the assumption that the dimension of the state of each node is
one \cite{yyliu, Lou, Zhang, Nabi, Wang14}.
% In such an over-simplified
%setting with single-dimensional edges, the classical controllability
%matrix-rank criterion can be directly applied to the entire network,
%which is judged by using the system controllability matrix just like
%for a single large-scale dynamical system investigated before (see
%textbooks, e.g., \cite{chui}).
However, most real-world networks of dynamical systems have
higher-dimensional node states, and many multi-input/multi-output
(MIMO) nodes are interconnected via multi-dimensional channels.
% in
%which from one node to another there are multiple and separate
%interactive communications in parallel.
% For notational convenience
%below, every node will be represented by its state vector when the
%coupling between two nodes is considered. Thus, for two connected
%nodes, which can have higher-dimensional state vectors, if their
%input and/or output are higher-dimensional (larger than one) then
%the nodes and their communication channel are referred to as MIMO,
%but if their input and output are both one-dimensional then their
%communication channel is single-input/single-output (SISO) while the
%node vectors can be either one-dimensional or higher-dimensional.
%For the MIMO setting, and also for the SISO setting with
%higher-dimensional node states, we will show that there are many
%unusual and counter-intuitive properties that are very different
%from the SISO setting with one-dimensional nodes, where the latter
%had been extensively investigated before as mentioned above. One
%case in point is Example \ref{exadd}, given below, which shows that
%a simple network of two mutually connected nodes with 2-dimensional
%state can be uncontrollable even if the two nodes are both
%controllable and observable individually. In fact, most
%aforementioned existing results obtained for the SISO setting with
%one-dimensional nodes {\it cannot\/} be directly extended to the
%general MIMO setting, as will be demonstrated below.
%Motivated by the above observations,
In this paper we study the controllability of networked
higher-dimensional systems with higher-dimensional nodes mainly for
the MIMO setting.

In the literature, few results are developed for the controllability
of networked higher-dimensional systems. Controllability and
observability of specified cartesian product networks are
investigated in \cite{2014tac}, while a general networked system is
considered in \cite{2015zt} with every subsystem subject to external
control input. It should be noted that some recent studies have
addressed consensus controllability of networks with MIMO nodes
\cite{Cai, Ni2013, Xiang}, where the controllability usually can be
decoupled into two independent parts: one is about the
controllability of each individual node and the other is solely
determined by the network topology. A general setting of complex
dynamical networks {\it cannot\/} be decoupled, however, as will be
seen from the discussions below.

This paper addresses networked MIMO LTI dynamical node-systems in a
directed and weighted topology, where there is no requirement for
every subsystem to have an external control input.
%Several typical network topologies are analyzed in
%detail, including chains, cycles, stars and trees, for which some
%specific conditions will be derived, as well as the setting of networked
%systems in a general topology.
Some controllability conditions on the network topology, node
dynamics, external control inputs and inner interactions are
developed, so that effective criteria can be obtained for
determining the large-scale networked system controllability.
 On one hand, some necessary and
sufficient conditions on the controllability of a networked system
with SISO higher-dimensional nodes are derived. On the other hand,
some interesting results on the controllability of a networked
system with MIMO nodes are obtained: 1) Under some mild conditions,
controllability and observability of the nodes together are
necessary for the controllability of the networked systems but,
nevertheless, they are not sufficient; 2) the controllability of the
network topology is necessary for the controllability of the
generated networked system; 3) the controllability of individual
node is necessary for the controllability of chain-networks, but not
necessary for cycle-networks; 4) interactions among the states of
different nodes play an important role in determining the
controllability of a general networked system. Surprisingly, for the
same network topology with the same node systems, the interactions
among the states of nodes not only can lead controllable nodes to
form an uncontrollable networked system, but also can assemble
uncontrollable nodes into a controllable networked system.

The rest of the paper is organized as follows: In Section
\ref{networked system}, some preliminaries and the general model of
networked MIMO LTI systems are presented. Controllability conditions
on various networked systems are investigated in Section \ref{main}.
Finally, conclusions are drawn with some discussions in Section
\ref{conclusion}.

\section{Preliminaries and the networked system}\label{networked
system}

\subsection{Preliminaries}\label{prelim}

Throughout, let $\mathbb{R}$ and $\mathbb{C}$ denote the real and
complex numbers respectively, $\mathbb{R}^n$ ($\mathbb{C}^n$) the
vector space of real (complex) $n$-vectors, $\mathbb{R}^{n\times m}$
($\mathbb{C}^{n\times m}$) the set of $n\times m$ real (complex)
matrices, $I_N$ the $N\times N$ identity matrix, and
$diag(a_1,\cdots,a_N)$ the $N\times N$ diagonal matrix with diagonal
elements $a_1,\cdots,a_N$. Denote by $\s(A)$ the set of all the
eigenvalues of matrix $A$ and by $\otimes$ the Kronecker product.

In a directed graph, an edge $(i,j)$ is directed from $i$ to $j$,
where $i$ is the tail and $j$ is the head of the edge. As reviewed
for a directed graph in \cite{yyliu}, a matching is a set of edges
that do not share any common tail or head, and a node being the head
of an edge in the matching is called a matched node; otherwise, it
is an unmatched node. A maximum matching is a matching that contains
the largest possible number of edges in the graph. A perfect
matching is a matching which matches all nodes in the graph. A graph
formed by a sequence of edges $\{(v_{i},
v_{i+1})\,|\,i=1,\cdots,\ell-1\}$ with no repeated node is called a
path, denoted as $v_1,\cdots, v_{\ell}$, where $v_1$ is the
beginning and $v_{\ell}$ is the end of the path, and $v_{\ell}$ is
said to be reachable from $v_1$. If $v_1,\cdots, v_{\ell}$ is a
path, then the graph formed by adding the edge $(v_{\ell},v_1)$ is a
cycle. A graph without cycles is called a tree. The node in a tree
which can reach every other node is called the root of the tree. A
leaf in a rooted tree is a node of degree 1 that is not the root.

Specifically, the notion of system controllability includes state
controllability and structural controllability. For an $n$-dimensional
system $\dot{x} = Ax+Bu$, it is said to be {\it state controllable},
if it can be driven from any initial state to the origin in finite
time by a piecewise continuous control input. $(A,B)$ is state
controllable if and only if the controllability matrix
$(B,AB,A^2B,\cdots,A^{n-1}B)$ has a full row rank
 \cite{Kalman, chui}. A parameterized system $(A,B)$ (i.e., all of their nonzero
elements are parameters) is said to be {\it structurally
controllable}, if it is possible to choose a set of nonzero
parameter values such that the resulting system $(A,B)$ is state
controllable \cite{Lin74}. In this paper, for brevity, {\it
controllability\/} always means state controllability unless
otherwise specified, e.g., structural controllability.

\subsection{The networked system model}\label{model}

Consider a general directed and weighted network consisting of
MIMO LTI node-systems in the following form:
 \bna\label{2.1}
 \left\{\begin{array}{lll}
 \dot{x}_i &=& Ax_i+\sum\limits_{j=1}^{N}\beta_{ij} Hy_j, \\
 y_i &=& C x_i, \qquad i=1,2,\cdots,N,
 \end{array}\right.
 \ena
in which $x_i\in \mathbb{R}^n$ is the state vector and $y_i\in
\mathbb{R}^m$ the output vector of node $i$, $H\in
\mathbb{R}^{n\times m}$ denotes the inner coupling matrix, and
$\b_{ij}\in \mathbb{R}$ represent the communication channels between
different nodes. As usual, assume that $\b_{ii}=0$ and $\b_{ij}\neq
0$ if there is an edge from node $j$ to node $i$, otherwise
$\b_{ij}= 0$, for all $i,j=1,2,\cdots,N$.

When subjected to control inputs, the above networked system becomes
 \bna\label{2.2}
 \dot{x}_i = Ax_i+\sum\limits_{j=1}^{N}\beta_{ij} HC
 x_j+\d_i Bu_i,\quad i=1,2,\cdots,N,
 \ena
where $u_i\in \mathbb{R}^p$ is the external control input to node
$i$, $B\in \mathbb{R}^{n\times p}$, with $\d_i=1$ if node $i$ is
under control, but otherwise $\d_i=0$, for all $i=1,2,\cdots,N$. To
avoid trivial situations, always assume that $N\geq 2$.

Here and throughout, for statement simplicity a network consisting
of more than one dynamical node, with or without control inputs
(e.g., \dref{2.1} and \dref{2.2}), will be called a {\it networked
system}.

Denote
 \bna\label{l}
 L=[\b_{ij}]\in \mathbb{R}^{N\times N} \quad{\rm and}\quad
 \Delta=diag(\d_1,\cdots,\d_N),
 \ena
which represent the network topology and the external input
channels of the networked system \dref{2.2}, respectively. Let
$X=[x_1^T,x_2^T,\cdots,x_N^T]^T$ be the whole state of the
networked system, and $U=[u_1^T,u_2^T,\cdots, u_N^T]^T$ the total
external control input. Then, this networked system can be
rewritten in a compact form as
 \bna\label{2.4}
 \dot{X}= \Phi X+ \Psi U\,,
 \ena
with
 \bna\label{2.5}
 \Phi =I_N\otimes A+L\otimes HC,\quad \Psi=\Delta \otimes B.
 \ena

In this paper, the focus is on how the network topology (described
by the matrix $L$), the node-system $(A,B,C)$, the external control
input (determined by the matrix $\Delta$), and the inner
interactions specified by $H$ affect the controllability of the
whole networked system.

\subsection{Some counter-intuitive examples}\label{examples}

In \cite{yyliu}, it is shown that a network is structurally
controllable if and only if there is an external input on each
unmatched node and there are directed paths from controlled nodes
with input signals to all matched nodes. For the networked system
\dref{2.4}-\dref{2.5} formed by nodes with higher-dimensional state
vectors, however, its controllability can be much more complicated,
as demonstrated by the following example.

\begin{example}\label{exadd2}
Consider a network of three identical nodes, with
$\b_{21}=\b_{31}=1$, and $\delta_1=1$, $\delta_2=\delta_3=0$. It is
not structurally controllable with one external input if each node
has a one-dimensional state, since nodes $2$ and $3$ can not be
matched simultaneously. If each node has a higher-dimensional state
with
 \bna\nn
 A=\left[\begin{array}{ll}
 a_{11} & a_{12}\\
 a_{21} & a_{22}
 \end{array}\right],
 B=\left[\begin{array}{ll}
 1 & 0\\
 0 & 1
 \end{array}\right],
 H=\left[\begin{array}{ll}
 h_1 & 0\\
 0 & h_2
 \end{array}\right],
 C=\left[\begin{array}{ll}
 1 & 0\\
 0 & 1
 \end{array}\right],
 \ena
where $a_{ij} \not= 0$ and $h_i \not= 0$, $i,j=1,2$, then obviously
$(A,B)$ is controllable and $(A,C)$ is observable. However, based on
the results from Subsection \ref{tree} below, one knows that the
networked system is uncontrollable for any matrix $A$, although it
is structurally controllable due to the existence of self-matched
cycle in every MIMO node.
\end{example}
%\begin{figure}[htbp]
%\centering
%  \subfigure[]{\begin{minipage}[c]{.1\textwidth}
%    \centering
%    \includegraphics[width=\textwidth]{fig1a.eps}
%  \end{minipage}}%
%   \hspace{.03\textwidth}
%  \subfigure[]{\begin{minipage}[c]{.3\textwidth}
%    \centering
%    \includegraphics[width=\textwidth]{fig1b.eps}
%  \end{minipage}}%
%  \caption{The network topology and the networked system.}
%   \label{addf}
%\end{figure}

The following three examples show that, even the network is a cycle
having a perfect matching, the controllability of $(A,B)$ is neither
necessary nor sufficient for the controllability of the whole
networked system.

\begin{example}\label{ex1}
Consider a network of two mutually connected identical nodes, with
$\b_{12}=\b_{21}=1$. Suppose that both nodes have external control
inputs, i.e. $\delta_1=\delta_2=1$, and
 \bna
 \nn
 A=\left[\begin{array}{ll}
 1 & 0 \\
 1 & 1
 \end{array}
 \right],
 B=\left[\begin{array}{l}
 1 \\
 0
 \end{array}
 \right],
 H=\left[\begin{array}{l}
 0 \\
 1
 \end{array}\right],
 C=\left[ 1 \ \ 0 \right].
 \ena
It is easy to check that $(A,B)$ is controllable. However, the
networked system
% \bna
% \nn
% \dot{X} =
% \left[\begin{array}{llll}
% 1 & 0&0&0 \\
% 1 & 1&1&0\\
% 0&0&1&0\\
% 1&0&1&1
% \end{array}\right] X
% +\left[\begin{array}{ll}
% 1 & 0 \\
% 0&0\\
% 0&1\\
% 0&0
% \end{array}\right]
% \left[\begin{array}{ll}
% u_1\\
% u_2
% \end{array}\right]
% \ena
\dref{2.2} is uncontrollable, although each node has an independent
external input.
\end{example}

\begin{example}\label{exadd}
Consider a simple network of two mutually connected identical nodes,
with $\b_{12}=\b_{21}=1$, $\delta_1=1$, $\delta_2=0$, and
 \bna
 \nn
 A=\left[\begin{array}{ll}
 1 & 0 \\
 1 & 1
 \end{array}\right],
 B=\left[\begin{array}{l}
 1 \\
 0
 \end{array}\right],
 H=\left[\begin{array}{l}
 0 \\
 1
 \end{array}\right],
 C=\left[ 0 \ \ 1 \right].
 \ena
Then, $(A,B)$ is controllable, $(A,C)$ is observable. However, the
networked system
% \bna
% \nn
% \dot{X} =
% \left[\begin{array}{llll}
% 1 & 0 & 0 & 0 \\
% 1 & 1 & 0 & 1 \\
% 0 & 0 & 1 & 0 \\
% 0 & 1 & 1 & 1
% \end{array}\right] X+
% \left[\begin{array}{l}
% 1 \\
% 0 \\
% 0 \\
% 0
% \end{array}\right]
% u_1
% \ena
\dref{2.2} is uncontrollable.
\end{example}

\begin{example}\label{ex2}
Consider a network of two mutually connected identical nodes, with
$\b_{12}=\b_{21}=1$, $\delta_1=1$, $\delta_2=0$, and
 \bna
 \nn
 A=\left[\begin{array}{ll}
 1 & 0 \\
 1 & 1
 \end{array}\right],
 B=\left[\begin{array}{l}
 0 \\
 1
 \end{array}\right],
 H=\left[\begin{array}{l}
 1 \\
 0
 \end{array}\right],
 C=\left[ 0 \ \ 1\right].
 \ena
Then, $(A,B)$ is uncontrollable. However, the networked system
 %\bna
% \nn
% \dot{X} =
% \left[\begin{array}{llll}
% 1 & 0 & 0 & 1 \\
% 1 & 1 & 0 & 0 \\
% 0 & 1 & 1 & 0 \\
% 0 & 0 & 1 & 1
% \end{array}\right] X+
% \left[\begin{array}{l}
% 0 \\
% 1 \\
% 0 \\
% 0
% \end{array}\right]
% u_1
% \ena
\dref{2.2} is controllable, although there is only one node under
external control.
\end{example}

Comparing the above three examples, their network topologies are the
same and their node-system matrices $A$ are identical. However,
these networked systems have very different controllabilities. The
interactions among the states of nodes not only can lead
controllable nodes to form an uncontrollable network, but also can
assemble uncontrollable nodes into a controllable network.

\section{Main Results}\label{main}

The controllability of the networked system \dref{2.4}-\dref{2.5} is
considered in this section, where results on general network
topologies are obtained, with specific and precise conditions
obtained for some typical networks including tress and cycles.

First, recall the Popov-Belevitch-Hautus (PBH) rank condition
\cite{Hautus}: the networked system \dref{2.4}-\dref{2.5} is
controllable if and only if
 \bna\label{pbh}
 \mbox{rank}(sI_{N\cdot n}-\Phi,\S) = N\cdot n
 \ena
is satisfied for any complex number $s$.

\subsection{A general network topology}\label{general}

Based on the PBH rank condition \dref{pbh}, one can prove the
following results.

\begin{thm}\label{t11}
If there exists one node without incoming edges, then to reach
controllability of the networked system \dref{2.4}-\dref{2.5}, it is
necessary that $(A,B)$ is controllable and moreover an external
control input is applied onto this node which has no incoming edges.
\end{thm}

\noindent\proof Assume that node $i$ does not have any incoming
edge. Then, the $i$th block row of $\Phi$ in \dref{2.5} becomes
$[0,\cdots,0,A,0,\cdots,0]$. If there is no external control input
onto node $i$, that is, $\d_i=0$, then for any $s_0\in \s(A)$, the
row rank of $[s_0I-\Phi,\Psi]$ will be reduced at least by 1. If
$(A,B)$ is uncontrollable, then there exists an $s_0\in \s(A)$ such
that $\mbox{rank}(s_0I-A,B)\leq n-1$, which will also result in the
reduction of the rank of $[s_0I-\Phi,\Psi]$.
\endproof

\begin{thm}\label{t12}
If there exists one node without external control inputs, then for
networked system \dref{2.4}-\dref{2.5} to be controllable, it is
necessary that $(A,HC)$ is controllable.
\end{thm}

\noindent\proof Assume that node $i$ does not have any external
input, that is, $\d_i=0$. If $(A,HC)$ is uncontrollable, then there
exist an $s_0\in \s(A)$ and a nonzero vector $\x\in
\mathbb{C}^{1\times n}$, such that $\x(s_0I-A)=0$ and $\x HC=0$. Let
$\a=[0,\cdots,0,\x,0 \cdots,0]$ with $\x$ located at the $i$th
block. Then, it is easy to verify that $\a(s_0I-\Phi)=0$ and
$\a\Psi=0$.
\endproof

\begin{thm}\label{t13}
If the number of nodes with external control inputs is $m$, and
$N>m\cdot\mbox{rank}(B)$, then for the networked system
\dref{2.4}-\dref{2.5} to be controllable, it is necessary that
$(A,C)$ is observable.
\end{thm}

\noindent\proof Suppose that $(A,C)$ is unobservable. Then, there
exist an $s_0\in \s(A)$ and a nonzero vector $\xi\in \mathbb{C}^n$
such that
 \bna\label{3.3}
 C \xi=0\quad{\rm and}\quad (s_0I-A)\xi=0\,.
 \ena

Consider the matrix $\P_{s_0}\triangleq s_0I-\P$, and partition it
into $N$ column blocks, $[\P_{s_0}^1,\P_{s_0}^2,\cdots,
\P_{s_0}^N]$, with
$\P_{s_0}^i=[\b_{1i}(HC)^T,\cdots,\b_{i-1,i}(HC)^T,(sI-A)^T,
\b_{i+1,i}$ $(HC)^T, \cdots,\b_{Ni}(HC)^T]^T$, which corresponds to
node $i$.

Based on formula \dref{3.3}, one has
 \bna
 \P_{s_0}^i\xi=0,\quad i=1,\cdots,N,
 \ena
which implies that $\mbox{rank}(\P_{s_0}^i)\leq n-1$. Therefore,
$\mbox{rank}(\P_{s_0})\leq N\cdot(n-1)$. In view of
$\mbox{rank}(\S)\leq m\cdot \mbox{rank}(B)<N$, one has
$\mbox{rank}(s_0I-\P,\S)< N\cdot n$, showing that the networked
system \dref{2.4}-\dref{2.5} is uncontrollable.
\endproof

%\noindent\proof Assume that there are $m$ nodes with external
%control inputs, labelled as node $1$ to node $m$. Then, $\d_i=1$ for
%$i=1,\cdots,m$, and $0$ for the others (which, therefore, could be
%removed from $\Psi$).
%
%For $s_0\in \s(A)$, there exists a nonzero vector $\x\in
%\mathbb{C}^{1\times n}$ such that $\x(s_0I-A)=0$. Choose
% \bna
% \nn
% \a_i&=&0\,, \quad\mbox{for}\quad i=1,\cdots,m, \\
% \nn
% \a_i&=&k_i\x\,, \quad\mbox{for}\quad i=m+1\cdots,N,
% \ena
%where $k_i\in \mathbb{R},$ $i=m+1,\cdots,N$, are undetermined
%parameters that satisfy
% \bna\label{3.18}
% \sum\limits_{i=m+1,i\neq j}^N k_{i}\b_{ij}=0, \qquad
% j=1,\cdots,N.
% \ena
%If node $j$ is a leaf node, then $\b_{ij}=0$ for $i=1,\cdots,N$,
%since a leaf node has no outgoing edge. Therefore, for a leaf node
%labelled as $j$, the $j$th equation in \dref{3.18} is satisfied for
%all $k_i$. Assume that there are $\ell$ ($\ell>m$) leaf nodes. Then,
%$N-\ell$ equations in \dref{3.18} should be satisfied by $N-m$
%parameters $k_i$. Hence, the dimension of the solution space is
%$\ell-m>0$, and so there are some nonzero solutions
%$[k_{m+1},\cdots,k_{N}]$ to equations \dref{3.18}. Choose any
%nonzero solution, and let $\a=[\a_1,\a_2,\cdots,\a_N]$ accordingly.
%Then, it is easy to verify that
% $$
% \a(s_0I-\P)=0\,,\quad \a\S=0\,,
% $$
%with $\P$ and $\S$ defined in \dref{2.5}. Therefore, the networked
%system \dref{2.4}-\dref{2.5} is uncontrollable.
%\endproof

\begin{thm}\label{t15}
If $(L,\Delta)$ is uncontrollable, then the networked system
\dref{2.4}-\dref{2.5} is uncontrollable.
\end{thm}

\noindent\proof If $(L,\Delta)$ is uncontrollable, then there exist
an $s_0\in \s(L)$ and a nonzero vector $\x \in \mathbb{C}^{1\times
N}$ such that
 \bna\label{3.20}
 \x(s_0I-L)=0\,,\quad \x \Delta=0\,.
 \ena
Therefore,
 \bna
 \nn
 (\x\otimes I)\dot{X} & =& (\x\otimes I)\big((I\otimes A
 +L\otimes HC)X + (\Delta\otimes B) U\big)\\
 \nn
 &= & (\x\otimes A + s_0 \x \otimes HC) X\\
 \nn
 &=& \big(\x \otimes (A+s_0 HC)\big)X\,,
 \ena
that is,
 \bna \label{3.20b}
 \left(\sum\limits_{i=1}^N \x_i x_i\right)^{\prime}
 =(A+s_0 HC)\sum\limits_{i=1}^N \x_i x_i\,.
 \ena
This implies that the variable $\sum\limits_{i=1}^N \x_i x_i$ is
unaffected by the external control input $U$. For the zero initial
state $x_i(t_0)=0$, $i=1,\cdots, N$, one has $\sum\limits_{i=1}^N
\x_i x_i(t_0)=0 $. Moreover, $\sum\limits_{i=1}^N \x_i x_i(t)=0$ for
all $t>t_0$, because of the uniqueness of the solution to the linear
equation \dref{3.20b}. Consequently, for any state $
\widetilde{X}\triangleq [\widetilde{x}_1^T, \cdots,
\widetilde{x}_N^T]^T$ with $\sum\limits_{i=1}^N \x_i \widetilde{x}_i
\neq 0$, there is no external control input $U$ that can drive the
networked system \dref{2.4}-\dref{2.5} to traverse from state $0$ to
$\widetilde{X}$. Thus, it is uncontrollable.
\endproof

If the network is not structurally controllable by external inputs,
then $(L,\Delta)$ is uncontrollable, and thus the networked system
will be uncontrollable. Since a network having more leaf nodes than
nodes with external control input is not structurally controllable
\cite{RR2014}, the following result comes accordingly.

\begin{corollary}\label{t14}
If there are more leaf nodes than the nodes with external control
inputs, then the networked system \dref{2.4}-\dref{2.5} is
uncontrollable.
\end{corollary}

If $(L,\Delta)$ is controllable, Examples \ref{ex4}, \ref{ex6}, and
\ref{ex8} discussed bellow for specific network topologies show
that, for the networked system \dref{2.4}-\dref{2.5} to be
controllable, it is not sufficient to ensure $(A,B)$ and $(A,HC)$ be
both controllable and $(A,C)$ be observable. Moreover, Example
\ref{ex6} shows that even every node has an external input and
$(A,B)$ is controllable, the networked system may still be
uncontrollable.

Next, some necessary and sufficient conditions for the
controllability of networked systems with SISO nodes are developed.
First, a lemma is given.

\begin{lemma}\label{t3}
Assume that $C\in \mathbb{R}^{1\times n}$ is nonzero. Then, $(A,HC)$
is controllable if and only if $(A,H)$ is controllable.
\end{lemma}

\noindent\proof Since $C\in \mathbb{R}^{1\times n}$, one has $H\in
\mathbb{R}^{n\times 1}$ and $\mbox{rank}(HC)=1$. Therefore,
$\mbox{rank}(sI-A,HC)=\mbox{rank}(sI-A,H)$, leading to the
conclusion.
\endproof

Before moving on to the theorem, some new notations are needed.
Denote the set of nodes with external control inputs by
 \bna\label{nu}
 \mathfrak{U} = \{i \,|\, \d_i\neq 0,\, i=1,\cdots,N\}\,.
 \ena
For any $s\in \s(A)$, define a matrix set
 \bna
 \Gamma(s)=\left\{\rsx{[\a_1^T,\cdots,\a_N^T]\;}\;
 \begin{array}{ll}
 \a_i \in
 \Gamma^1(s)\;\mbox{for}\; i\notin \mathfrak{U}\\
 \a_i \in \Gamma^2(s)\;\mbox{for}\; i\in \mathfrak{U}
 \end{array}\right\},
 \ena
where
 \bna
 \nn
 \Gamma^1(s)&=&\{\x\in \mathbb{C}^{1\times n}\;|\;\x(sI-A)=0\},\\
 \nn \Gamma^2(s)&=&\{\x\in \mathbb{C}^{1\times n}\;|\;\x B=0,\;
 \x\in \Gamma^1(s)\}.
 \ena

\begin{thm}\label{t18}
Suppose that $|\mathfrak{U}|<N$, $B\in \mathbb{R}^{n\times 1}$, and
$C\in \mathbb{R}^{1\times n}$. Then, the networked system
\dref{2.4}-\dref{2.5} is controllable if and only if the following
hold:
\begin{enumerate}[(i)]
 \item $(A,H)$ is controllable;
 \item $(A,C)$ is observable;
 \item for any $s\in \s(A)$ and $\kappa \in \Gamma(s)$,
       $\kappa L\neq0$ if $\kappa \neq 0$;
 \item for any $s\notin \s(A)$,
       $\mbox{rank}(I-L\gamma,\Delta\eta)=N$,
 with $\g=C(sI-A)^{-1}H$ and $\eta=C(sI-A)^{-1}B$.
 \end{enumerate}
\end{thm}

\noindent\proof {\it Necessity}. From Theorem \ref{t12} and Lemma
\ref{t3}, it follows that condition {\it (i)} is necessary. From
Theorem \ref{t13}, it follows that condition {\it (ii)} is also
necessary.

Now, suppose that condition {\it (iii)} is not necessary. Then,
there exist an $s_0\in \s(A)$ and a nonzero matrix $\kappa \in
\Gamma(s_0)$ such that
 $$
\kappa L=0\,.
 $$
For matrix $M \in \mathbb{C}^{p \times q}$, denote by $vec(M)\in
\mathbb{C}^{pq \times 1}$ the vectorization of matrix $M$ formed by
stacking the columns of $M$ into a single column vector.
Furthermore, let $\a=vec(\kappa)^T$. Since $\kappa \in \Gamma(s_0)$,
it is easy to verify that $\a \Psi=0$ and
 \bna
 \nn
 \a(s_0I-\Phi)&=&\a(I_N\otimes (s_0I-A)-L\otimes HC)\\
 \nn
 &=& -\a (L\otimes HC)\\
 \nn
 &=& -vec(C^T H^T \kappa L)^T=0\,,
 \ena
which contradicts the network controllability.

Finally, suppose that condition {\it (iv)} is not necessary. Then,
there exists an $s_0 \notin \s(A)$ satisfying
 $$
 \mbox{rank}(I-L\gamma_0,\Delta\eta_0) < N\,,
 $$
with $\g_0=C(s_0I-A)^{-1}H$ and $\eta_0=C(s_0I-A)^{-1}B$. Thus,
there exists a nonzero vector $\zeta=[\zeta_1,\cdots,\zeta_N]\in
\mathbb{C}^{1\times N}$, such that
 $$
\zeta (I-L\gamma_0)=0 \quad\mbox{and}\quad \zeta \Delta\eta_0=0\,.
 $$
Let $\a=[\a_1,\cdots,\a_N]$ with $\a_i=\zeta_i C(s_oI-A)^{-1}$.
Then, since $\zeta \neq 0$, one has $\a\neq 0$. Moreover,
 \bna
 \nn  \a\Psi
 &=&(\zeta\otimes C(s_0I-A)^{-1})\cdot(\Delta\otimes B)\\
 \nn  &=& (\zeta \Delta) \otimes  (C(s_oI-A)^{-1}B)\\
 \nn &=& \zeta \Delta \eta_0 =0\,,
 \ena
and
 \bna\nn &&\a(s_0I-\Phi)\\
 \nn &=&(\zeta\otimes C(s_oI-A)^{-1})\cdot(I_N\otimes
 (s_0I-A)-L\otimes HC)\\\nn &=& \zeta \otimes C -
 \zeta L \otimes (C(s_oI-A)^{-1}H)C\\
 \nn &=& (\zeta-\zeta L\gamma_0)\otimes C =0\,.
 \ena
This is also in conflict with the controllability of the networked
system.

\noindent{\it Sufficiency}. For $s\in \mathbb{C}$, suppose that
there exists a vector $\a=[\a_1,\cdots,\a_N]$, with $\a_i \in
\mathbb{C}^{1\times n}$, such that $\a(sI-\Phi)=0$ and $\a\Psi=0$.
That is,
 \bna\label{a3.1}
 \a_i(sI-A)-\sum\limits_{j\neq i}\b_{ji}\a_j HC=0,\ \ i=1\cdots,N,
 \ena
and
 \bna\label{a3.2}
 \a_i B=0\,,\quad i\in \mathfrak{U}\,.
 \ena

If $s\in \s(A)$, then $\mbox{rank}(sI-A)<n$. From \dref{a3.1}, it
follows that, for all $i=1,\cdots,N$,
 \bna\label{a3.3}
 \sum\limits_{j\neq i}\b_{ji}\a_j H=0\,.
 \ena
If not, then
 $\mbox{rank}\left(\left[\begin{array}{c}
 C\\
 sI-A
\end{array}\right]\right)=\mbox{rank}(sI-A) < n$,
which contradicts with the observability of $(A,C)$. Moreover, based
on \dref{a3.1}, one has
 \bna\label{a3.5}
 \a_i(sI-A)=0\,,\quad i=1\cdots, N\,.
 \ena
Therefore, for all $i=1\cdots, N$, one has
 \bna\label{a3.4}
 \sum\limits_{j\neq i}\b_{ji}\a_j(sI-A)=0\,.
 \ena
Combining it with \dref{a3.3} and the controllability of $(A,H)$,
one obtains
 \bna\label{a3.6}
 \sum\limits_{j\neq i}\b_{ji}\a_j=0\,,\quad i=1\cdots, N.
 \ena

Next, let $\kappa=[\a_1^T,\cdots,\a_N^T]$. In view of \dref{a3.2},
\dref{a3.5} and \dref{a3.6}, it is easy to verify that $\kappa L=0$
with $\a_i(sI-A)=0$ for $i=1,\cdots,N$, and $\a_i B=0$ for $i\in
\mathfrak{U}$. Therefore, by condition {\it (iii)}, one has $\a=0$.

If $s\notin \s(A)$, then $sI-A$ is invertible. From \dref{a3.1}, one
has
 \bna\label{a3.7}
 \a_i=\sum\limits_{j\neq i}\b_{ji}\a_j HC(sI-A)^{-1},\quad
 i =1,\cdots, N.
 \ena
Let $\zeta_i=\sum\limits_{j\neq i}\b_{ji}\a_jH$. Then, for
$i=1,\cdots,N$,
 \bna\label{a3.8}
 \a_i=\zeta_iC(sI-A)^{-1}\,,
 \ena
and
 \bna\label{a3.9}
 \begin{array}{lll} \zeta_i&=&\sum\limits_{j\neq i}\b_{ji}\a_j
 H = \sum\limits_{j\neq i}\b_{ji}\zeta_jC(sI-A)^{-1}H \\
 &=& \sum\limits_{j\neq i}\b_{ji}\zeta_j \gamma\,.
 \end{array}
 \ena
Let $\zeta=[\zeta_1,\cdots,\zeta_N]$, and rewrite \dref{a3.9} as
 \bna\label{a3.10}
 \zeta(I-L\gamma)=0\,.
 \ena
Then, from \dref{a3.2} and \dref{a3.8}, it follows that
$\zeta_iC(sI-A)^{-1}B=0$ for $i\in \mathfrak{U}$, which is
equivalent to
 \bna\label{a3.11}
 \zeta\Delta\eta=0\,.
 \ena
Consequently, by combining it with \dref{a3.10} and condition {\it
(iv)}, one has $\zeta=0$, which together with \dref{a3.8} imply that
$\a=0$.

It follows from the above analysis that, for any $s\in \mathbb{C}$,
the row vectors of matrix $[sI-\Phi,\Psi]$ are linearly independent,
hence $\mbox{rank}(sI-\Phi,\Psi)=N\cdot n$. Thus, the networked
system \dref{2.4}-\dref{2.5} is controllable.
\endproof

Next, some typical network structures, i.e., trees and circles are
discussed in details.

\subsection{Trees} \label{tree}

In view of Corollary \ref{t14}, the following result comes easily.

\begin{corollary}
\label{t10} Consider a tree-network, in which every node is
reachable from the root, and only the root has an external control
input. If there is more than one leaf node in the tree, then the
networked system is uncontrollable. Consequently, a star networked
system with $N>2$ is uncontrollable.
\end{corollary}

A tree with only one leaf is a chain, which could be described by a
path $1\rightarrow 2\rightarrow\cdots\rightarrow n$. Based on
Theorem \ref{t11}, node $1$ should be under external control.

This chain-networked system \dref{2.4}-\dref{2.5} has
 \bna\label{3.2}
 \Phi=\left[
 \begin{array}{cccc}
 A&0&\cdots&0\\
 \b_{21}HC&A&&\\
 \vdots&\ddots&&\vdots\\
 0&\cdots&\b_{N,N-1}HC &A
 \end{array}\right],\quad
 \Psi=\left[
 \begin{array}{l}
 B\\
 0\\
 \vdots\\
 0
 \end{array}\right],
 \ena
where $\b_{i,i-1}\neq 0$ for $i=2,\cdots,N$, and $\b_{ij}=0$ for
$j\neq i-1$, $i=1,\cdots,N$.

From Theorems \ref{t11} and \ref{t12}, one obtains the following
result.

\begin{corollary}\label{t1}
A necessary condition for the controllability of the chain networked
system \dref{2.4}-\dref{3.2} is that $(A,B)$ and $(A,HC)$ are both
controllable.
\end{corollary}

\begin{remark}\label{rm1}
The observability of $(A,C)$ is not necessary for the
controllability of the chain networked system \dref{2.4}-\dref{3.2},
as shown by the following example.
\end{remark}

\begin{example}\label{ex3}
Consider a chain-network of two identical nodes, with $\b_{21}=1$
and
 \bna
 \nn
 A=\left[\begin{array}{cc}
 1 & 0 \\
 1 & 1
 \end{array}\right],\;
 B=\left[\begin{array}{cc}
 1&0 \\
 0&1
 \end{array}\right],\;
 H=\left[\begin{array}{c}
 1\\
 0
 \end{array}\right],\;
 C=\left[ 1 \ \ 0 \right].
 \ena

It is easy to check that $(A,B)$ and $(A,HC)$ are both controllable
and $(A,C)$ is unobservable. The coupled networked system \dref{3.2}
has $\mbox{rank}(\S,\P\S,\P^2\S,\P^{3}\S)=4$,  indicating that it is
controllable.
 %\bna
% \Phi=\left[
% \begin{array}{llll}
% 1&0&0&0\\
% 1&1&0&0\\
% 1&0&1&0\\
% 0&0&1&1\\
% \end{array}\right],\quad
% \Psi=\left[
% \begin{array}{ll}
% 1&0\\
% 0&1\\
% 0&0\\
% 0&0
% \end{array}\right],
% \ena
%which is controllable.
Therefore, the observability of $(A,C)$ is
indeed not necessary.
\end{example}

\begin{remark}\label{rm2}
Suppose that $(A,B)$ and $(A,HC)$ are both controllable and $(A,C)$
is observable. However, these are not sufficient to guarantee the
controllability of the chain networked system \dref{2.4}-\dref{3.2},
as shown by the following example.
\end{remark}

\begin{example}\label{ex4}
Consider a chain-network of two nodes, with $\b_{21}=1$ and
 \bna
 \begin{array}{lllll}
 \nn
 A=\left[\begin{array}{cc}
 1 & 2 \\
 5 & 4
 \end{array}\right],\;
 B=\left[\begin{array}{c}
 2 \\
 -1
 \end{array}
 \right],
 H=\left[\begin{array}{cc}
 -1&1\\
 -4&1
 \end{array}\right],\;
 C=\left[\begin{array}{cc}
 1 & 0\\
 0 & 1
 \end{array}\right].
 \end{array}
 \ena

It is easy to check that $(A,B)$ and $(A,HC)$ are both controllable
and $(A,C)$ is observable. However, the coupled networked system
\dref{3.2}
% \bna
% \Phi=\left[
% \begin{array}{cccc}
% 1&2&0&0\\
% 5&4&0&0\\
% -1&1&1&2\\
% -4&1&5&4\\
% \end{array}\right],\quad
% \Psi=\left[\begin{array}{cc}
% 2\\
% -1\\
% 0\\
% 0
% \end{array}\right]
% \ena
has $\mbox{rank}(6\cdot I-\Phi,\S)=3<4$, indicating that it is
uncontrollable.
\end{example}

If the input and output channels are all one-dimensional, namely, if
all the nodes are SISO, then Theorem \ref{t18} can be restated as
follows.

\begin{corollary}\label{t4}
Assume that $B\in \mathbb{R}^{n\times 1}$ and $C\in
\mathbb{R}^{1\times n}$. The chain networked system
\dref{2.4}-\dref{3.2} is controllable if and only if $(A,B)$ and
$(A,H)$ are both controllable and $(A,C)$ is observable.
\end{corollary}

\proof  Construct $\kappa=[\a_1^T,\cdots,\a_N^T]$ according to
condition {\it (iii)} of Theorem \ref{t18}, such that $\a_1\in
\Gamma^2$ satisfies $\a_1(sI-A)=0$ and $\a_1 B=0$, and moreover
$\a_i \in \Gamma^1$ satisfies $\a_i(sI-A)=0$ for $i=2,\cdots,N$. In
view of $\kappa L=
[\b_{21}\a_2^T,\b_{32}\a_3^T,\cdots,\b_{N,N-1}\a_N^T,0]$, the
condition $\kappa L\neq 0$ for $\kappa \neq 0$ is equivalent to
$\a_1=0$, which implies the equivalence with the controllability of
$(A,B)$.  Therefore, condition {\it (iii)} in Theorem \ref{t18} is
equivalent to the controllability of $(A,B)$.

 Condition
{\it(iv)} in Theorem \ref{t18} is automatically satisfied for the
chain-network, since
 \bna\label{a3.12}
 L=\left[
 \begin{array}{cccc}
 0&0&\cdots&0\\
 \b_{21}&0&&\\
 \vdots&\ddots&&\vdots\\
 0&\cdots&\b_{N,N-1} &0
 \end{array}\right],
 \ena
and correspondingly for $s\notin \s(A)$, $\mbox{rank} (I-L\gamma)=N$
with $\g=C(sI-A)^{-1}H$.
\endproof

\subsection{Cycles} \label{sub2}

Now, assume that the network topology is a cycle. Since the cycle
has a perfect matching, one external input is enough for the
structural controllability, which can be added to any node in the
cycle. Without loss of generality, assume that node 1 is under
external control.

The cycle networked system has
 \bna\label{3.7}
 \begin{array}{ll}
 \Phi=\left[
 \begin{array}{cccc}
 A&0&\cdots&\b_{1N}HC\\
 \b_{21}HC&A&&\\
 \vdots&\ddots&&\vdots\\
 0&\cdots&\b_{N,N-1}HC &A
 \end{array}\right],\\
 \Psi=\left[
 \begin{array}{cccc}
 B^T&0&\cdots&0
 \end{array} \right]^T,
 \end{array}
 \ena
where $\b_{1N}\neq 0$, $\b_{i,i-1}\neq 0$ for $i=2,\cdots,N$, and
$\b_{ij}=0$ otherwise.

From Theorem \ref{t12}, the controllability of $(A,HC)$ is necessary
for the controllability of the networked system
\dref{2.4}-\dref{3.7}.

\begin{remark}\label{rm3}
The controllability of $(A,B)$ and the observability of $(A,C)$ are
not necessary for the controllability of the cycle networked system
\dref{2.4}-\dref{3.7}, as can be seen from the following example.
\end{remark}

%Example \ref{ex2} has shown that the controllability of $(A,B)$ is
%not necessary for the controllability of the cycle-network. The
%following example further shows that the observability of $(A,C)$ is
%not necessary for the controllability of the cycle-network either.

\begin{example}\label{ex5}
Consider a cycle-network of three identical nodes, with
$\b_{13}=\b_{21}=\b_{32}=1$ and
 \bna
 \begin{array}{lllll}
 \nn
 A=\left[\begin{array}{llll}
 0&1&0&0 \\
 0&0&1&0\\
 0&0&0&1\\
 0&0&0&1
 \end{array} \right],\;
 B=\left[\begin{array}{lll}
 1&0&0 \\
 0&1&0\\
 0&0&1\\
 0&0&0
 \end{array}\right],\;
 H=\left[\begin{array}{ll}
 0&1\\
 0&0\\
 0&0\\
 1&0
 \end{array}
 \right],\\
 C=\left[\begin{array}{llll} 0 & 1&0&0\\
 0&0&1&0
 \end{array}\right].
 \end{array}
 \ena
It is easy to check that $(A,B)$ is uncontrollable and $(A,C)$ is
unobservable. However, the coupled system \dref{3.7} has
$\mbox{rank}(\S,\P\S,\P^2\S,\cdots,\P^{11}\S)=12$, indicating that
the networked system is controllable.
\end{example}

\begin{remark}\label{rm4}
Conditions that $(A,B)$ and $(A,HC)$ are both controllable and
$(A,C)$ is observable together are not sufficient to guarantee the
controllability of the cycle networked system \dref{2.4}-\dref{3.7},
as shown by the following example.
\end{remark}

\begin{example}\label{ex6}
Consider a cycle-network of three identical nodes, with
$\b_{13}=\b_{21}=\b_{32}=1$ and
 \bna
 \begin{array}{lllll}
 \nn
 A=\left[\begin{array}{ll}
 1 & 1 \\
 0 &1
 \end{array}\right],\;
 B=\left[\begin{array}{l}
 1 \\
 1
 \end{array}\right],\;
 H=\left[\begin{array}{ll}
 0&0\\
 0&1
 \end{array}\right],\;
 C=\left[\begin{array}{ll} 1&0\\
 0&1
 \end{array}\right].
 \end{array}
 \ena
It is easy to check that $(A,B)$ and $(A,HC)$ are both controllable
and $(A,C)$ is observable. However, although every node is driven by
a control input, the whole networked system with
 \bna
 \Phi=\left[
 \begin{array}{lll}
 ~A&~~0&HC\\
 HC&~A&~~0\\
 ~~0&HC&~A\\
 \end{array}\right],\quad
 \Psi=\left[
 \begin{array}{lll}
 B&0&0\\
 0&B&0\\
 0&0&B
 \end{array}\right]
 \ena
has $\mbox{rank}(\S,\P\S,\P^2\S,\cdots,\P^{5}\S)=5<6$, implying that
the networked system is uncontrollable.
\end{example}

\begin{remark} \label{rm5}
Even every node is SISO, the controllability of $(A,B)$ is not
necessary for the controllability of the networked system
\dref{2.4}-\dref{3.7}, as shown by the following example.
\end{remark}

\begin{example}\label{ex7}
Consider a cycle-network of three identical nodes, with
$\b_{13}=\b_{21}=\b_{32}=1$ and
 \bna
 \begin{array}{lllll}
 \nn
 A=\left[\begin{array}{ll}
 0 & 1 \\
 0 &0
 \end{array}\right],\;
 B=\left[\begin{array}{l}
 1 \\
 0
 \end{array}\right],\;
 H=\left[\begin{array}{l}
 0\\
 1
 \end{array} \right],\;
 C=\left[1\ \ 0 \right].
 \end{array}
 \ena
It is easy to check that $(A,H)$ is controllable, $(A,C)$ is
observable, and $(A,B)$ is uncontrollable. However, the
 coupled system \dref{3.7}
has $\mbox{rank}(\S,\P\S,\P^2\S,\cdots,\P^{5}\S)=6$, indicating that
the networked system is controllable.
\end{example}

\begin{remark}\label{rm6}
Assume that every node is SISO. The conditions that $(A,B)$ and
$(A,H)$ are controllable and $(A,C)$ is observable together are not
sufficient to guarantee the controllability of the networked system
\dref{2.4}-\dref{3.7}, as shown by the following example.
\end{remark}

\begin{example}\label{ex8}
Consider a cycle-network of three identical nodes, with
$\b_{13}=-1$, $\b_{21}=\b_{32}=1$, and
 \bna
 \begin{array}{lllll}
 \nn
 A=\left[\begin{array}{lll}
 1 & 8 & 7 \\
 4 & 5 & 6 \\
 1 & 2 & 3
\end{array}\right],
 B=\left[\begin{array}{l}
 1 \\
 0\\
 1
 \end{array}\right],
 H=\left[\begin{array}{l}
 1\\
 1\\
 1
 \end{array}\right],
C=\left[  4\ \ 3\ \ 6 \right].
 \end{array}
 \ena
It is easy to check that $(A,B)$ and $(A,H)$ are both controllable
and $(A,C)$ is observable. However, the coupled system \dref{3.7}
has $\mbox{rank}(\S,\P\S,\P^2\S,\cdots,\P^{8}\S)=8<9$, showing that
the networked system is uncontrollable.
\end{example}

For circle network with SISO nodes, a new criterion for the
controllability is given as follows.

\begin{thm}\label{t8}
Assume that $B\in \mathbb{R}^{n\times 1}$ and $C\in
\mathbb{R}^{1\times n}$. The cycle networked system
\dref{2.4}-\dref{3.7} is controllable if and only if $(A,H)$ is
controllable, $(A,C)$ is observable, and moreover
 \bna\label{3.8}
 \mbox{rank}\left(I-bHC(sI-A)^{-1},B\right)=n\,,
 \quad \forall s \notin \s(A)\,,
 \ena
where $b=\b_{1N}\prod\limits_{i=1}^{N-1}\b_{i+1,i}\gamma^{N-1}$,
with $\g=C(sI-A)^{-1}H$.
\end{thm}

\proof For the cycle-network,
 \bna\label{a3.13}
 L=\left[
 \begin{array}{cccc}
 0&0&\cdots&\b_{1,N}\\
 \b_{21}&0&&\\
 \vdots&\ddots&&\vdots\\
 0&\cdots&\b_{N,N-1} &0
 \end{array}\right],
 \ena
which is invertible, therefore condition {\it(iii)} in Theorem
\ref{t18} is automatically satisfied.

In the following, it will be proved that condition {\it(iv)} in
Theorem \ref{t18} is equivalent to the above rank condition. Note
that the two conditions are both given in terms of matrix ranks, yet
one is about the network topology which is $N$-dimensional while the
other is about a sub-system which is only $n$-dimensional.

If $\g=0$, then the two matrices both have full ranks. In the
following, assume that $\g\neq 0$.

 If $\mbox{rank}(I-L\gamma,
\Delta\eta)<N$, then there exists a nonzero vector
$\mathbf{k}=[k_1,\cdots,k_N]\in \mathbb{C}^{1\times N}$ such that
 $$
 \mathbf{k}=\mathbf{k}L\gamma
 \quad\mbox{and}\quad\mathbf{k}\Delta\eta=0\,,
 $$
that is,
 \bna\label{a3.14}
 \begin{array}{lll}
 k_i&=&k_{i+1}\b_{i+1,i}\g,\quad i=1,\cdots, N-1,\\
 k_N&=& k_1\b_{1,N}\g, \\
 k_1 \eta &=& 0\,.
 \end{array}
 \ena
From the recursion formula \dref{a3.14}, it follows that $k_1\neq 0$
since $\mathbf{k}\neq 0$. Moreover, $k_1=k_2\b_{21}\g=\cdots
=k_N\prod\limits_{i=1}^{N-1}\b_{i+1,i}\g^{N-1}=k_1b\g$, which
implies that $b\g=1$. Choose $\xi=k_1C(sI-A)^{-1}$. Then, $\xi\neq
0$, $\xi B=k_1C(sI-A)^{-1}B =k_1 \eta=0$, and $\xi(I-bHC(sI-A)^{-1})
=k_1C(sI-A)^{-1}-k_1b\g C(sI-A)^{-1}=0$, which implies that
$\mbox{rank}(I-bHC(sI-A)^{-1},B)<n$.

If  $\mbox{rank}(I-bHC(sI-A)^{-1},B)<n$, then there exists a nonzero
vector $\xi\in \mathbb{C}^{1\times n}$, satisfying
 $$
 \xi=b\xi HC(sI-A)^{-1}, \quad \xi B=0\,.
 $$
Since $\xi\neq 0$, one has $b\neq 0$ and $\xi H\neq 0$.  Moreover,
$\xi H=b\xi H \g$, which implies that $b\g=1$. Now, define
 \bna\label{a3.15}
 \begin{array}{lll}
 k_1&=& b\xi H,\\
 k_N&=&\b_{1,N}k_1 \g, \\
 k_i&=& k_{i+1}\b_{i+1,i}\g\,, \quad i=2,\cdots,N-1.
 \end{array}
 \ena
One can easily verify that
 $$
 k_1 \eta= b\xi H C(sI-A)^{-1}B=\xi B=0\,,
 $$
 $$
 k_2\b_{21}\g=k_N\prod\limits_{i=1}^{N-1}\b_{i+1,i}\g^{N-1}
 =k_1b\g=k_1\,.
 $$
Therefore, $\mathbf{k}=\mathbf{k}L\gamma$ and $\mathbf{k}
\Delta\eta=0$ with $\mathbf{k}=[k_1,\cdots,k_N]$, which implies that
$\mbox{rank}(I-L\gamma,\Delta\eta)<N$.
\endproof

Looking back to Example \ref{ex7}, it can be seen that
$\s(A)=\{0,0\}$. And, for any $s\neq 0$, one has $b=s^{-4}$ and
 \bna
 \nn&&\mbox{rank}\left(I-bHC(sI-A)^{-1},B\right)\\
 \nn&=&\mbox{rank}
 \left(\left[
 \begin{array}{ll}
 ~1&~~0\\
 -s^{-5}&1-s^{-6}
 \end{array}
 \right],\; \left[
 \begin{array}{l}
 1\\
 0
 \end{array}
 \right]\right)=2.
 \ena
Moreover, $(A,H)$ is controllable and $(A,C)$ is observable.
Therefore, from Theorem \ref{t8}, it follows that the networked
system in Example \ref{ex7} is controllable.

Looking back to Example \ref{ex8}, it can be seen that for
$s=2\notin \s(A)$, one has $C(2I-A)^{-1}H=-1$, $b=-1$, and
 \bna
 \nn&&\mbox{rank}(I-bHC(2I-A)^{-1},B)\\
 \nn&=&\mbox{rank}
 \left(\left[
 \begin{array}{lll}
 1&-1&0\\
 0&~~0&0\\
 0&-1&1
 \end{array}
 \right],\; \left[
 \begin{array}{l}
 1\\
 0\\
 1
 \end{array}
 \right]\right)=2<3.
 \ena
Therefore, from Theorem \ref{t8}, it follows that the networked
system in Example \ref{ex8} is uncontrollable.

\section{Conclusions}\label{conclusion}

We have investigated a network consisting of MIMO LTI node-systems
$(A,B,C)$, in a topology described by matrix $L$ with inner
interactions described by matrix $H$, with or without control inputs
determined by matrix $\Delta$. We have studied the integrated
effects of the network topology $L$, node-system $(A,B,C)$ and inner
interactions $H$ on the controllability of the networked system.

We have shown that a networked system in the MIMO setting is
uncontrollable if the network topology $L$ is uncontrollable by
external inputs through $\Delta$, e.g., a non-trivial star-network
with a single input to its root. For a networked system to be
controllable, the controllability of $(A,B)$ and $(A,HC)$, as well
as the observability of $(A,C)$, are necessary under some
conditions; but they are not sufficient in general, even for the
cycle-network which has a perfect matching.

For SISO nodes with higher-dimensional state vectors, we have
presented necessary and sufficient conditions for the
controllability of some networked systems, including trees, cycles
as well as a general network topology. These results not only
provide precise and efficient criteria for determining the
controllability of large-scale networked systems, by means of
verifying some properties of a few matrices of lower dimensions, but
also provide some general guidelines on how to assemble
uncontrollable nodes to form a controllable networked system, which
is deemed useful in engineering practice.

If each node-system (described by higher-dimensional matrices
$(A,B,H,C)$) is viewed as a sub-network, then the networked system
studied in this paper can also be considered as an interdependent
network (or interconnected network, multi-layer network, network of
networks, multiplex network, etc. \cite{Bocca2014}); therefore, the
results obtained in this paper should shed lights onto studying the
controllability of such complex networks.

\begin{ack}
The authors thank Professor Tong Zhou from Tsinghua University,
Beijing for some helpful discussions. This work was supported by the
National Natural Science Foundation of China under Grant Nos.
61374176 and 61473189, and the Science Fund for Creative Research
Groups of the National Natural Science Foundation of China (No.
61221003), as well as the Huawei Technologies Co. Ltd.
\end{ack}

%\appendix
%\section{A summary of Latin grammar}    % Each appendix must have a short title.
%\section{Some Latin vocabulary}         % Sections and subsections are supported
%                                        % in the appendices.
\end{document}